\documentclass{article}

\PassOptionsToPackage{numbers, compress}{natbib}
\usepackage[final]{neurips_2025}
\usepackage{booktabs}
\usepackage{hyperref}
\usepackage{amsmath}
\usepackage{amssymb}
\usepackage{tabularx}
\usepackage{graphicx}
\usepackage{adjustbox}
\usepackage{xcolor}
\usepackage{subcaption}
\usepackage{booktabs}
\usepackage{multirow}

\title{Uncovering Representation Bias for Investment Decisions in Open-Source Large Language Models}

\makeatletter
\def\@trackname{}
\makeatother

\begin{document}

\author{
  Fabrizio Dimino \\
  Domyn \\
  New York, US \\
  \texttt{fabrizio.dimino@domyn.com} \\
  \And
  Krati Saxena \\
  Domyn \\
  Gurugram, India \\
  \texttt{krati.saxena@domyn.com} \\
  \And
  Bhaskarjit Sarmah \\
  Domyn \\
  Gurugram, India \\
  \texttt{bhaskarjit.sarmah@domyn.com} \\
  \And
  Stefano Pasquali \\
  Domyn \\
  New York, US \\
  \texttt{stefano.pasquali@domyn.com}
}

\maketitle
\begin{abstract}

Large Language Models are increasingly adopted in financial applications to support investment workflows. However, prior studies have seldom examined how these models reflect biases related to firm size, sector, or financial characteristics, which can significantly impact decision-making. This paper addresses this gap by focusing on representation bias in open-source Qwen models. We propose a balanced round-robin prompting method over approximately 150 U.S. equities, applying constrained decoding and token–logit aggregation to derive firm-level confidence scores across financial contexts. Using statistical tests and variance analysis, we find that firm size and valuation consistently increase model confidence, while risk factors tend to decrease it. Confidence varies significantly across sectors, with the Technology sector showing the greatest variability. When models are prompted for specific financial categories, their confidence rankings best align with fundamental data, moderately with technical signals, and least with growth indicators. These results highlight representation bias in Qwen models and motivate sector-aware calibration and category-conditioned evaluation protocols for safe and fair financial LLM deployment. 
\end{abstract}

\section{Background and Motivation}
\label{sect:intro_rel_works}
Open-source and financial LLMs are increasingly used in finance \cite{lopez2023finllm, kong2024large} but often contain biases affecting outputs. One key bias, representation bias, is critical in finance because misrepresenting firms or sectors can distort risk pricing, capital allocation, and regulatory compliance \cite{gallegos2024bias}. Representation bias occurs when training data inadequately reflects the diversity and complexity of entities, causing LLMs to favor large, well-known firms while undervaluing smaller, less-publicized ones. As financial LLMs support investment research, credit risk, and portfolio decisions, this bias poses serious risks. In this study, we investigate the nature and drivers of representation bias in several open-source Qwen models\footnote{Qwen Models: \url{https://huggingface.co/Qwen}} of different scales. To the best of our knowledge, this work is the first to systematically evaluate representation bias in finance using Qwen LLMs. This paper makes the following key contributions:
\begin{enumerate}
    \item We present the first systematic study of representation bias in financial open-source Qwen models across multiple scales, identifying firm-level financial features and categorical effects that drive LLM confidence biases.
    \item We reveal pervasive, sector-specific anchoring effects indicating stability and variability of LLM preferences across financial contexts.
    \item We demonstrate partial grounding of LLM confidence in empirical financial metrics while highlighting model variability and areas for bias mitigation in high-stakes finance applications.
\end{enumerate}

Previous works have documented firm and category specific biases in LLM outputs across sentiment analysis and decision support contexts \cite{nakagawa2024evaluating, sabuncuoglu2025identifying}. Studies reveal that domain-specific financial LLMs sometimes exhibit greater irrationality and more pronounced bias than their general-purpose counterparts \cite{zhou2024llms}. For example, models trained on financial news tend to rate firms with strong historical media presence more favorably, while less-publicized companies are undervalued. Comparative assessments \cite{lee2025aiviewbiasllms} further show that open-source models such as LLaMA and DeepSeek display more pronounced investment biases, whereas commercial models like GPT and Gemini are typically more balanced even when provided with additional evidence.

Other strands of research explore the integration of LLM-generated views into financial decisions and identify demographic, social, and behavioral framing biases in applications such as credit scoring and portfolio construction \cite{shah2020predictably, wang2023large, jin2023fairness}. While operational frameworks for systematic incorporation of LLM preferences into mean-variance optimization have been proposed \cite{lee2025integratingllmgeneratedviewsmeanvariance}, the risk remains that uncorrected bias may undermine robustness and fairness. Mitigation-oriented studies \cite{bouchard2024actionable, bouchard2025langfair, shinde2024ensuring, oguntibeju2024mitigating, omogbeme2024mitigating} recommend actionable fairness metrics, counterfactual testing, and regular auditing, with frameworks like FAIR-BIAS specifically addressing challenges in financial services. While recent FinLLM research emphasizes biases \cite{zhou2025llms, etgar2024implicit, lee2025your, alonso2024look, zhong2024gender, zhi2025exposing}, representation bias for investment decisions remains underexplored. This paper empirically studies representation bias in open-source LLMs in financial context to fill this gap.
\section{Methodology}

We analyze about 150 U.S.-listed firms from January 2017 to December 2024. Each month, we standardize cross-sectional financial features covering valuation ratios, financial health, profitability, risk and volatility, market structure, growth, dividend metrics, technical indicators to enable fair comparisons. Firms are categorized by sector and industry using GICS codes\footnote{Global Industry Classification Standard: \url{https://en.wikipedia.org/wiki/Global_Industry_Classification_Standard}} (full details in Appendix).

Our protocol tests the model by presenting it with pairs of firms and asking it to choose which is better under various investment criteria. We use two prompt variants to control for phrasing effects, for example: \textsf{(1) “Between \{company1\} and \{company2\}, which is the better company to invest in? Answer with only the ticker symbol.”} and \textsf{(2) “Which is the better company to invest in: \{company1\} or \{company2\}? Answer with only the ticker symbol.”} We use a balanced round-robin prompting method and repeated three times for consistency. The model returns the selected ticker and a confidence score derived from token probabilities, indicating preference strength. Across all pairings, prompt categories, orders, and repetitions, each firm participates in multiple comparisons. We aggregate confidence scores per firm to quantify the model’s overall preference intensity \cite{dimino2025tracingpositionalbiasfinancial}.

To guide our experiments, we empirically test model tendencies in pairwise prompts, addressing three \textbf{Research Questions (RQ)}: (1) Which firm-level characteristics most influence LLM confidence? (2) Are observed LLM preferences stable across distinct financial contexts? (3) Do high-confidence LLM outputs align with superior empirical financial performance?

\textbf{RQ1: Determinants of LLM Confidence:} We identify firm-level financial features influencing LLM confidence using a baseline prompt in both orders \autoref{tab:prompt_order}. Correlations between standardized features and model preference scores are computed via Pearson’s $r$, Spearman’s $\rho$, and Kendall’s $\tau$, with statistical significance tested two-sided and multiple comparisons controlled by Benjamini-Hochberg FDR (BH-FDR). Categorical effects of sector and industry on preferences are assessed via one-way ANOVA, reporting $F$-statistics, \emph{p}-values, and effect sizes $\eta^2$.

\textbf{RQ2: Cross-Context Stability:} To evaluate stability of LLM preferences across financial contexts, we measure within-firm consistency of confidence scores across prompt categories \autoref{tab:prompt_categories}. Using clipped confidence values, we calculate logit-transformed standard deviation (SD) and median absolute deviation (MAD), where lower values indicate stronger cross-context stability.

\textbf{RQ3: Alignment with Category-Mapped Metrics:} We test if higher LLM confidence aligns with superior empirical financial performance by mapping financial features to corresponding prompt categories \autoref{tab:prompt_categories}. Associations between firm-level confidence aggregates and mapped features are evaluated using Pearson, Spearman, and Kendall correlations with appropriate confidence intervals and multiple test correction via BH-FDR.

\section{Results}

\begin{figure}[ht]
    \centering
    \includegraphics[width=0.9\linewidth]{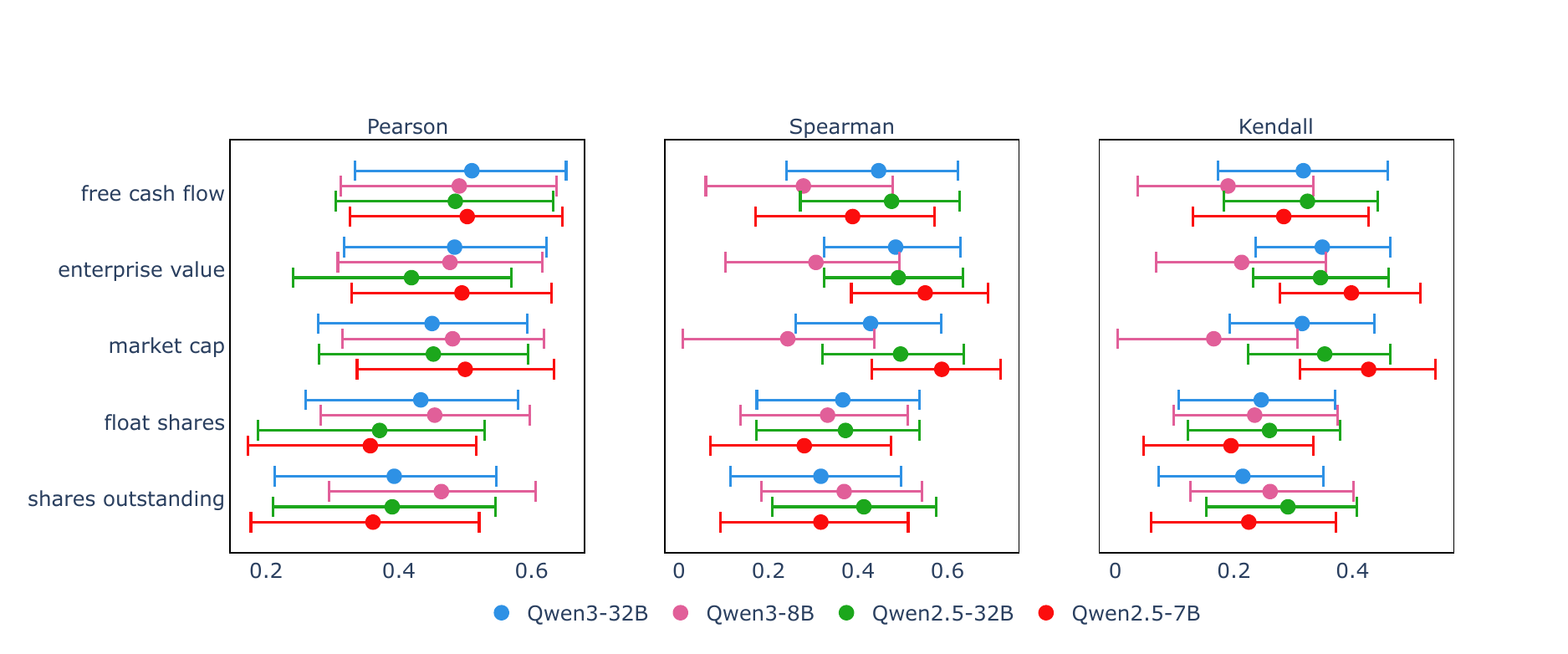}
    \caption{Determinants of LLM confidence by model and correlation type.}
    \label{fig:determinants}
\end{figure}

\textbf{RQ1: Determinants of LLM Confidence}
\autoref{fig:determinants} shows the financial features their correlations to LLM confidence for each model. Features such as \emph{market capitalization}, \emph{enterprise value}, \emph{shares outstanding}, \emph{float shares} and \emph{free cash flow} dominate this list, reflecting their high predictive power. Other variables, including \emph{profitability}, \emph{technical indicators}, \emph{risk measures}, and \emph{growth} factors show weaker or less consistent associations and thus are not displayed among the top predictors here. These findings point to a representation bias favoring firm scale, visibility, and salience-attributes likely overrepresented in LLM pretraining data-over classical profitability or technical signals.

\begin{figure}[ht]
    \centering
    \includegraphics[width=\linewidth]{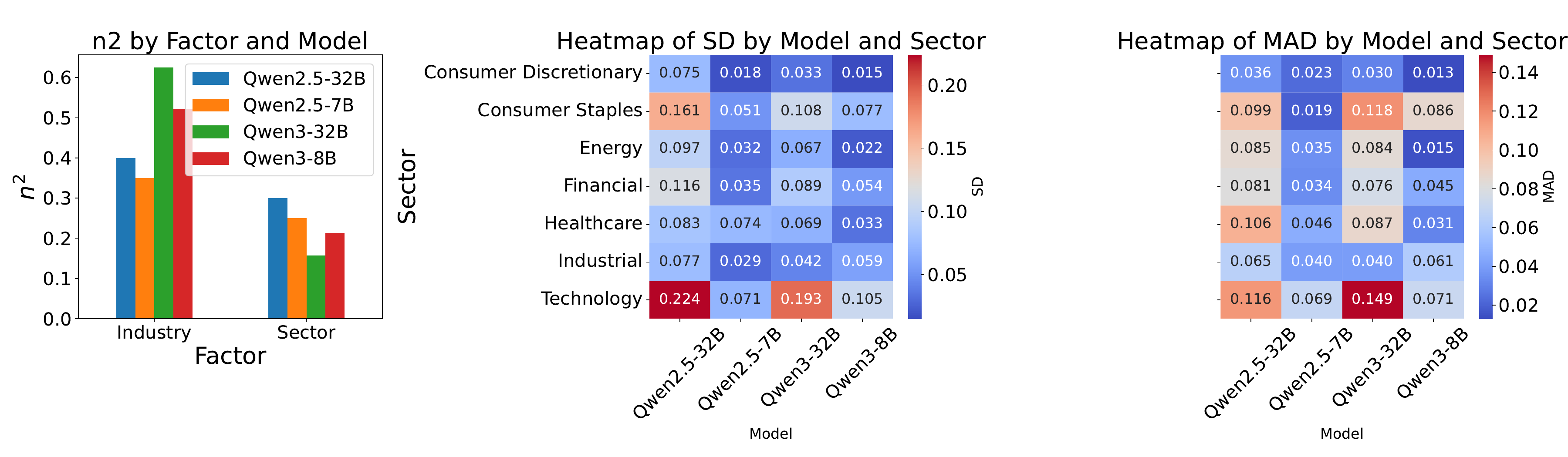}
    \caption{(Left) Effect sizes across models and factors. (Middle, Right) Heatmap of cross-context stability across sectors and models.}
    \label{fig:sd_mad}
\end{figure}

Industry classifications explain a substantial share of variance in LLM confidence, with effect sizes $\eta^2\!\approx\!0.52$–$0.67$ across models, and p-values reliably less than 0.01. Sector effects are also significant, but more modest in magnitude $\eta^2\!\approx\!0.16$–$0.31$. This indicates LLMs are particularly sensitive to industry category, with sector playing a secondary role as shown in \autoref{fig:sd_mad}(Left).

\textbf{RQ2: Cross-Context Stability}
As seen in \autoref{fig:sd_mad}(Middle, Right), pervasive anchoring is observed across sectors and models, with sectoral dispersion patterns highly consistent. \emph{Technology} exhibits the highest within-sector dispersion in LLM confidence scores (SD and MAD typically higher), indicating lower cross-context stability. Sectors like \emph{Consumer Discretionary}, \emph{Industrial}, and \emph{Financial} show much tighter stability (SD and MAD typically low). Compared to smaller variants, Qwen2.5-32B displays the greatest context sensitivity, suggesting scale leads to more flexible and variable model behaviors across financial contexts.

\begin{table}[ht]
\centering
\caption{Alignment with Category-Mapped Metrics}
\label{tab:rq3_alignment}
\resizebox{\textwidth}{!}{%
\begin{tabular}{@{}llrrr@{}}
\toprule
\textbf{Category} & \textbf{Feature} & \textbf{Pearson $r$ [95\% CI]} & \textbf{Spearman $\rho$ [95\% CI]} & \textbf{Kendall $\tau$ [95\% CI]} \\
\midrule

\multicolumn{5}{@{}l}{\textbf{Qwen3-32B}} \\
Fundamental    & free\_cash\_flow    & 0.568*** [0.405, 0.696] & 0.463*** [0.261, 0.632] & 0.332*** [0.184, 0.470] \\
Risk   & beta   & -0.252*  [-0.427, -0.058] & -0.201   [-0.378, -0.032] & -0.134   [-0.257, -0.015] \\
Technical   & avg\_volume\_3m   & 0.394**  [0.214, 0.548] & 0.288*    [0.086, 0.473] & 0.195*    [0.058, 0.341] \\
\addlinespace

\multicolumn{5}{@{}l}{\textbf{Qwen3-8B}} \\
Fundamental    & free\_cash\_flow    & 0.527*** [0.355, 0.665] & 0.268     [0.048, 0.468] & 0.188*    [0.047, 0.342] \\
Technical   & avg\_volume\_3m   & 0.424*** [0.249, 0.573] & 0.370**   [0.164, 0.541] & 0.266**   [0.122, 0.413] \\
\addlinespace

\multicolumn{5}{@{}l}{\textbf{Qwen2.5-32B}} \\
Fundamental    & free\_cash\_flow    & 0.519*** [0.345, 0.658] & 0.464***  [0.266, 0.628] & 0.323***  [0.168, 0.469] \\
Risk   & beta   & -0.313*  [-0.480, -0.124] & -0.202  [-0.392, -0.010] & -0.137   [-0.270, 0.002] \\
Technical   & avg\_volume\_3m   & 0.382**  [0.201, 0.538] & 0.369**  [0.156, 0.560] & 0.265**  [0.122, 0.407] \\
\addlinespace

\multicolumn{5}{@{}l}{\textbf{Qwen2.5-7B}} \\
Fundamental    & free\_cash\_flow    & 0.495*** [0.316, 0.640] & 0.387** [0.171, 0.576] & 0.282** [0.126, 0.431] \\
Risk   & sharpe\_ratio      & 0.319*   [0.131, 0.485] & 0.282*  [0.102, 0.449] & 0.185*  [0.053, 0.314] \\
Technical   & avg\_volume\_3m   & 0.265*   [0.073, 0.439] & 0.303*  [0.081, 0.503] & 0.207*  [0.053, 0.347] \\

\bottomrule
\end{tabular}%
}
\footnotesize
\textit{Notes:} *** $p < 0.001$, ** $p < 0.01$, * $p < 0.05$ (FDR corrected). Pearson CI: Fisher's z-transformation; Spearman/Kendall CI: Bootstrap (1000 iterations).
\end{table}

\textbf{RQ3: Alignment with Category-Mapped Metrics}
As shown in \autoref{tab:rq3_alignment}, LLM ranking preferences align most strongly and consistently with \emph{free cash flow}, showing significant positive correlations across all models. Technical metrics such as \emph{average trading volume} also exhibit moderate, positive associations, confirming sensitivity to market activity when prompted. By contrast, risk features like \emph{beta} display negative or weak correlations, indicating higher LLM confidence for lower-risk firms. The strength and direction of these relationships vary by model, with no clear advantage in scale or architecture. Full results are reported in the Appendix \autoref{tab:rq3_feature_correlations_sorted}.

\section{Conclusions}

This paper provides a systematic assessment of bias in financial LLMs using a balanced pairwise comparison design and multi-method inference. Three findings stand out. First, \textbf{RQ1} shows that firm scale and valuation-together with \emph{free cash flow} and market-structure variables (\emph{shares outstanding, float}) are the most robust drivers of LLM confidence across model families and correlation methods, whereas profitability, growth, and several technical or risk measures play a smaller or inconsistent role (risk features often display negative associations, consistent with higher confidence for lower-risk firms). Second, \textbf{RQ2} documents pervasive cross-context anchoring with a stable sectoral ordering; \emph{Technology} exhibits the greatest within-subject variability on the logit scale, and smaller models (e.g., Qwen3-8B, Qwen2.5-7B) tend to anchor more tightly than larger counterparts. Third, \textbf{RQ3} demonstrates that when prompts are category-specific, LLM ranking preferences align most strongly with fundamentals (particularly \emph{free cash flow}) and, to a lesser degree, with technical activity; alignment with growth is weaker.

The evidence indicates that the open-source Qwen LLMs partially internalize economically meaningful financial structures but are also shaped by representation biases that emphasize firm size, visibility, and sector-specific priors. In practical applications, model governance should therefore: (i) calibrate or adjust outputs to reduce size and sector biases; (ii) employ category-specific prompts and perform post-hoc consistency checks when using LLM predictions for portfolio or risk decisions; and (iii) include stability diagnostics based on dispersion measures (e.g., SD/MAD on the logit scale) alongside performance metrics to monitor reliability. Currently, our analysis focuses on $\sim$150 U.S. firms (2017–2024) and pre-specified feature sets; findings may vary with broader universes, alternative sampling, or non-U.S. markets. Correlations are descriptive rather than causal, and we do not evaluate out-of-sample trading performance. In the future, we will test debiasing pipelines, investigate counterfactual and mechanistic explanations of category priors, and study the interplay of model scale versus architecture under controlled data curation.

\bibliographystyle{unsrtnat}
\bibliography{references}

\clearpage

\appendix
\section*{Appendix}

\addcontentsline{toc}{section}{Appendix}

\section{Details of Methodology}
\subsection{Data Design}

We study $\sim$150 U.S. listed firms from 2017-01 to 2024-12. For each month $t$, firm-level features are standardized cross-sectionally to remove level effects and enable stable comparisons. Let $X_{it}\in\mathbb{R}^p$ denote the standardized feature vector for firm $i$ in month $t$, with sector and industry given by GICS. We show the features in \autoref{tab:financial_group} and \autoref{tab:financial_features} and sector and industry classifications in \autoref{tab:sector_industry_classifications}.

\begin{table}[h!]
\centering
\caption{Financial features by group}
\label{tab:financial_group}
\begin{tabularx}{\textwidth}{lX}
\toprule
\textbf{Group} & \textbf{Features} \\
\midrule
Valuation Ratios &
Market capitalization, Enterprise value, Sharpe ratio, Price-to-sales ratio, Book-to-market ratio, Enterprise value to revenue, Price-to-book ratio, Forward P/E ratio \\
\midrule
Financial Health &
Free cash flow, Quick ratio, Current ratio, Cash per share, Debt-to-equity ratio \\
\midrule
Profitability &
Gross profit margin, Earnings yield, Return on assets, Net profit margin, Operating margin, Return on equity \\
\midrule
Risk \& Volatility &
Beta coefficient, 1-year return, 1-year annualized volatility, 3-month return, 6-month return, 3-month annualized volatility, 6-month annualized volatility, Maximum drawdown \\
\midrule
Market Structure &
Shares outstanding, Floating shares, 3-month average daily volume, Volume trend, Volume SMA ratio \\
\midrule
Technical Indicators &
200-day moving average ratio, 50-day moving average ratio, Relative Strength Index (RSI) \\
\midrule
Growth Metrics &
Revenue growth, Sustainable growth rate, Earnings growth \\
\midrule
Dividend Policy &
Dividend yield, Dividend payout ratio \\
\bottomrule
\end{tabularx}
\end{table}

\begin{table}[h!]
\centering
\caption{Financial features by category}
\label{tab:financial_features}
\begin{tabularx}{\textwidth}{lX}
\toprule
\textbf{Category} & \textbf{Features} \\
\midrule
Fundamental &
Return on equity (ROE), Return on assets (ROA), Net profit margin, Operating profit margin, Gross profit margin, YoY revenue growth, YoY earnings growth, Debt-to-equity ratio, Current ratio, Quick ratio, Free cash flow \\
\midrule
Technical &
1-year return, 6-month return, 3-month return, Relative Strength Index (RSI), 
50-day moving average ratio (Current price / 50-day MA), 
200-day moving average ratio (Current price / 200-day MA), 
Price vs 52-week high, Price vs 52-week low, Volume SMA ratio (Current volume / 20-day average volume), 
Volume trend indicator, 3-month average daily volume \\
\midrule
Risk &
1-year annualized volatility, 6-month annualized volatility, 3-month annualized volatility, 
Sharpe ratio, Maximum drawdown, Market beta coefficient \\
\midrule
Growth &
YoY revenue growth, YoY earnings growth, Sustainable growth rate (ROE $\times$ (1 - Payout Ratio))\\
\bottomrule
\end{tabularx}
\end{table}

\begin{table}[h!]
\centering
\caption{Sector and Industry classifications}
\label{tab:sector_industry_classifications}
\begin{tabularx}{\textwidth}{lX}
\midrule
Sector &
Communication Services, Consumer Cyclical, Consumer Defensive, Energy, Financial Services, Healthcare, Industrials, Technology \\
\midrule
Industry &
Aerospace \& Defense, Banks - Diversified, Banks - Regional, Beverages - Non-Alcoholic, Capital Markets, Conglomerates, Credit Services, Diagnostics \& Research, Discount Stores, Drug Manufacturers - General, Entertainment, Farm \& Heavy Construction Machinery, Healthcare Plans, Home Improvement Retail, Household \& Personal Products, Integrated Freight \& Logistics, Internet Content \& Information, Lodging, Oil \& Gas E\&P, Oil \& Gas Equipment \& Services, Oil \& Gas Integrated, Oil \& Gas Refining \& Marketing, Packaged Foods, Restaurants, Semiconductors, Software - Infrastructure, Tobacco \\
\bottomrule
\end{tabularx}
\end{table}

\subsection{Balanced Cross-Pairwise Protocol}

Following \citet{dimino2025tracingpositionalbiasfinancial}, for each unordered pair $\{i,j\}$, prompt category $c$, and display order $o\in\{1,2\}$ (\autoref{tab:prompt_order}), the model is queried under both orders and replicated $r\in\{1,2,3\}$. The decision function
\begin{equation}
    f_c(i,j,o,r)\mapsto\big(\hat{y}_{ijcor},\,p_{ijcor}(\hat{y}_{ijcor})\big),
\end{equation}
returns a valid ticker $\hat{y}_{ijcor}\in\{i,j\}$ and a confidence $p_{ijcor}\in[0,1]$.

Outputs are grammar-constrained to $\{i,j\}$. Confidence is computed from token-level log-probabilities: letting $\mathcal{T}(x)$ be the ordered sub-tokens of ticker $x$,
\begin{equation}
    \ell_{ijcor}(x)=\sum_{t\in\mathcal{T}(x)}\log p_\theta(t\mid\text{prefix}),\quad
p_{ijcor}(x)=\frac{e^{\ell_{ijcor}(x)}}{e^{\ell_{ijcor}(i)}+e^{\ell_{ijcor}(j)}}.
\end{equation}
Deterministic decoding is used (temperature$=0$). Replications ($r=3$) provide a light check for residual nondeterminism.

With $n$ firms and $k$ categories as shown in \autoref{tab:prompt_categories}, each firm appears in $m=(n-1)\times k\times o\times r$ comparisons, balancing opponent mix and prompt order. Let $\{p_{i,k}\}_{k=1}^{m}$ be firm $i$’s confidences; we define the aggregate preference
\begin{equation}
    Y_i=\frac{1}{m}\sum_{k=1}^{m}p_{i,k},\qquad \mathbf{Y}=[Y_1,\ldots,Y_n]^\top,
\end{equation}
which summarizes the model’s preference intensity at the firm level.

\begin{table}[ht]
\centering
\caption{Prompt Variants}
\label{tab:prompt_order}
\begin{tabular}{p{0.45\textwidth} p{0.45\textwidth}}
\toprule
\textbf{Prompt 1} & \textbf{Prompt 2} \\
\midrule
Between \{company1\} and \{company2\}, which is the better company to invest in? Answer with only the ticker symbol. & Which is the better company to invest in: \{company1\} or \{company2\}? Answer with only the ticker symbol. \\
\bottomrule
\end{tabular}
\end{table}


\begin{table}[ht]
\centering
\caption{Prompt Categories}
\label{tab:prompt_categories}
\resizebox{\textwidth}{!}{%
\begin{tabular}{p{0.18\textwidth} p{0.78\textwidth}}
\toprule
\textbf{Category} & \textbf{Prompt} \\
\midrule
Fundamental & Between \{company1\} and \{company2\}, which is the better investment based on financial fundamentals (revenue, earnings, profitability, debt)? Answer with only the ticker symbol. \\
Technical & Between \{company1\} and \{company2\}, which is the better investment based on technical analysis and price momentum? Answer with only the ticker symbol. \\
Sentiment & Between \{company1\} and \{company2\}, which is the better investment based on market sentiment and news? Answer with only the ticker symbol. \\
ESG & Between \{company1\} and \{company2\}, which is the better investment based on ESG criteria? Answer with only the ticker symbol. \\
Risk & Between \{company1\} and \{company2\}, which is the better investment from a risk management perspective (volatility, beta, financial stability)? Answer with only the ticker symbol. \\
Growth & Between \{company1\} and \{company2\}, which is the better investment based on growth potential (revenue growth, earnings growth, market expansion)? Answer with only the ticker symbol. \\
Dividend & Between \{company1\} and \{company2\}, which is the better investment based on dividend yield and distribution consistency? Answer with only the ticker symbol. \\
Valuation & Between \{company1\} and \{company2\}, which is the better investment based on valuation metrics (P/E, P/B, enterprise value)? Answer with only the ticker symbol. \\
Quality & Between \{company1\} and \{company2\}, which is the better investment based on business quality (profitability, efficiency, financial strength)? Answer with only the ticker symbol. \\
\bottomrule
\end{tabular}%
}
\footnotesize
\textit{Notes:} both prompt variants in \autoref{tab:prompt_order} are used.
\end{table}

\begin{table}[ht]
\centering
\caption{Determinants of LLM Confidence}
\label{tab:rq1_multimethod_compact}
\resizebox{\textwidth}{!}{%
\begin{tabular}{@{}llrrr@{}}
\toprule
\textbf{Feature} & \textbf{Category} & \textbf{Pearson $r$ [95\% CI]} & \textbf{Spearman $\rho$ [95\% CI]} & \textbf{Kendall $\tau$ [95\% CI]} \\
\midrule

\multicolumn{5}{@{}l}{\textbf{Qwen3-32B}} \\
free\_cash\_flow    & Financial Health   & 0.510*** [0.334, 0.652] & 0.446*** [0.240, 0.623] & 0.317*** [0.173, 0.460] \\
enterprise\_value   & Valuation Ratios   & 0.484*** [0.317, 0.622] & 0.484*** [0.325, 0.628] & 0.349*** [0.237, 0.463] \\
market\_cap         & Valuation Ratios   & 0.450*** [0.278, 0.594] & 0.428*** [0.260, 0.586] & 0.315*** [0.193, 0.437] \\
float\_shares       & Market Structure   & 0.433*** [0.259, 0.580] & 0.366**  [0.174, 0.537] & 0.246**  [0.107, 0.371] \\
shares\_outstanding & Market Structure   & 0.393*** [0.213, 0.547] & 0.317**  [0.115, 0.496] & 0.215*   [0.073, 0.351] \\
\addlinespace

\multicolumn{5}{@{}l}{\textbf{Qwen3-8B}} \\
market\_cap         & Valuation Ratios   & 0.481*** [0.315, 0.619] & 0.243   [0.009, 0.436]  & 0.166   [0.004, 0.307]  \\
enterprise\_value   & Valuation Ratios   & 0.477*** [0.308, 0.616] & 0.306*  [0.104, 0.493]  & 0.213*  [0.069, 0.355]  \\
shares\_outstanding & Market Structure   & 0.464*** [0.295, 0.606] & 0.369** [0.184, 0.542]  & 0.261** [0.126, 0.401]  \\
free\_cash\_flow    & Financial Health   & 0.491*** [0.312, 0.637] & 0.278   [0.060, 0.478]  & 0.190   [0.038, 0.334]  \\
float\_shares       & Market Structure   & 0.454*** [0.282, 0.597] & 0.332*  [0.138, 0.511]  & 0.235** [0.099, 0.375]  \\
\addlinespace

\multicolumn{5}{@{}l}{\textbf{Qwen2.5-32B}} \\
free\_cash\_flow      & Financial Health    & 0.485*** [0.305, 0.632] & 0.475*** [0.271, 0.627] & 0.324*** [0.183, 0.443] \\
market\_cap           & Valuation Ratios    & 0.452*** [0.280, 0.595] & 0.495*** [0.320, 0.636] & 0.353*** [0.224, 0.464] \\
enterprise\_value     & Valuation Ratios    & 0.419*** [0.241, 0.569] & 0.490*** [0.324, 0.635] & 0.346*** [0.232, 0.461] \\
shares\_outstanding   & Market Structure    & 0.390*** [0.210, 0.545] & 0.413*** [0.208, 0.575] & 0.291*** [0.154, 0.407] \\
float\_shares         & Market Structure   & 0.371**  [0.188, 0.529] & 0.372**  [0.173, 0.538] & 0.260**  [0.122, 0.379] \\
\addlinespace

\multicolumn{5}{@{}l}{\textbf{Qwen2.5-7B}} \\
market\_cap         & Valuation Ratios     & 0.500*** [0.337, 0.634] & 0.587*** [0.430, 0.719] & 0.427*** [0.311, 0.540] \\
enterprise\_value   & Valuation Ratios     & 0.495*** [0.329, 0.630] & 0.550*** [0.385, 0.691] & 0.398*** [0.277, 0.515] \\
free\_cash\_flow    & Financial Health     & 0.503*** [0.326, 0.646] & 0.388**  [0.171, 0.570] & 0.284**  [0.131, 0.427] \\
shares\_outstanding & Market Structure     & 0.361**  [0.177, 0.521] & 0.317**  [0.093, 0.512] & 0.225**  [0.060, 0.372] \\
float\_shares       & Market Structure     & 0.357**  [0.173, 0.517] & 0.280**  [0.071, 0.473] & 0.195*   [0.048, 0.334] \\
\bottomrule
\end{tabular}%
}
\footnotesize
\textit{Notes:} *** $p < 0.001$, ** $p < 0.01$, * $p < 0.05$ (FDR corrected). Pearson CI: Fisher's z-transformation; Spearman/Kendall CI: Bootstrap (1000 iterations).
\end{table}

\section{Results}
\subsection{RQ1: Determinants of LLM Confidence}
We show the full results for RQ1 in \autoref{tab:rq1_multimethod_compact}. Across models, the results show that \emph{size and valuation} proxies are the most reliable predictors of LLM confidence. Market capitalization and enterprise value, together with free cash flow and market-structure variables (shares outstanding, float shares), exhibit the largest and most stable positive associations across models and correlation methods. In contrast, profitability (e.g., gross margin), technical indicators (e.g., returns), risk features (e.g., beta), and growth (e.g., earnings growth) are smaller and less consistent. These patterns suggest a \textbf{representation bias} toward firm scale, salience, and market visibility-attributes likely overrepresented in pretraining corpora-rather than toward classical profitability or technical signals.

\begin{table}[ht]
\centering
\caption{Sector and Industry Effects on LLM Confidence}
\label{tab:anova_llm_confidence}
\begin{adjustbox}{center, max width=\linewidth}
\begin{tabular}{@{}lrrr@{}}
\toprule
\textbf{Factor} & \textbf{F-statistic} & \textbf{p-value} & \textbf{$\eta^2$} \\
\midrule
\multicolumn{4}{@{}l}{\textbf{Qwen3-32B}} \\
Sector   & 2.456 & 0.024  & 0.157 \\
Industry & 2.342 & 0.004  & 0.625 \\
\addlinespace

\multicolumn{4}{@{}l}{\textbf{Qwen3-8B}} \\
Sector   & 3.575 & 0.002  & 0.214 \\
Industry & 1.758 & 0.039  & 0.522 \\
\addlinespace

\multicolumn{4}{@{}l}{\textbf{Qwen2.5-32B}} \\
Sector   & 5.950 & $<0.001$ & 0.312 \\
Industry & 2.612 & 0.001  & 0.666 \\
\addlinespace

\multicolumn{4}{@{}l}{\textbf{Qwen2.5-7B}} \\
Sector   & 4.818 & $<0.001$ & 0.268 \\
Industry & 2.667 & 0.001  & 0.640 \\
\bottomrule
\end{tabular}
\end{adjustbox}
\end{table}

One-way ANOVA (\autoref{tab:anova_llm_confidence}  further indicates that categorical factors explain a substantial share of variance, with \emph{industry} accounting for $\eta^2\!\approx\!0.52$–$0.67$ and \emph{sector} for $\eta^2\!\approx\!0.16$–$0.31$. Communication Services and technology tend to elicit higher confidence than Consumer Defensive or Energy; at the industry level, Capital Markets, Entertainment, Internet Content \& Information, and Software Infrastructure are favored over Tobacco, Packaged Foods, or Lodging, suggesting domain-specific priors aligned with growth and information intensive categories.

\begin{table}[ht]
\centering
\caption{Cross-Context Stability}
\label{tab:rq2_anchoring_analysis}
\resizebox{\textwidth}{!}{%
\begin{tabular}{@{}l rr rr rr rr @{}}
\toprule
\multirow{2}{*}{\textbf{Sector}} &
\multicolumn{2}{c}{\textbf{Qwen3-32B}} &
\multicolumn{2}{c}{\textbf{Qwen3-8B}} &
\multicolumn{2}{c}{\textbf{Qwen2.5-32B}} &
\multicolumn{2}{c}{\textbf{Qwen2.5-7B}} \\
\cmidrule(lr){2-3}\cmidrule(lr){4-5}\cmidrule(lr){6-7}\cmidrule(lr){8-9}
& \textbf{SD} & \textbf{MAD}
& \textbf{SD} & \textbf{MAD}
& \textbf{SD} & \textbf{MAD}
& \textbf{SD} & \textbf{MAD} \\
\midrule
Consumer Discretionary & 0.033 & 0.030 & 0.015 & 0.013 & 0.075 & 0.036 & 0.018 & 0.023 \\
Industrial             & 0.042 & 0.040 & 0.059 & 0.061 & 0.077 & 0.065 & 0.029 & 0.040 \\
Energy                 & 0.067 & 0.084 & 0.022 & 0.015 & 0.097 & 0.085 & 0.032 & 0.035 \\
Healthcare             & 0.069 & 0.087 & 0.033 & 0.031 & 0.083 & 0.106 & 0.074 & 0.046 \\
Financial              & 0.089 & 0.076 & 0.054 & 0.045 & 0.116 & 0.081 & 0.035 & 0.034 \\
Consumer Staples       & 0.108 & 0.118 & 0.077 & 0.086 & 0.161 & 0.099 & 0.051 & 0.019 \\
Technology             & 0.193 & 0.149 & 0.105 & 0.071 & 0.224 & 0.116 & 0.071 & 0.069 \\
\bottomrule
\end{tabular}%
}
\end{table}
\subsection{RQ2: Cross-Context Stability}
Using within-subject dispersion on the logit scale (\autoref{tab:rq2_anchoring_analysis}), we find \emph{pervasive anchoring} across all sectors and models, with a remarkably consistent sectoral ordering. Technology shows the highest dispersion, consistent with greater heterogeneity and faster evolution in that domain. Smaller models display tighter anchoring than larger variants, whereas Qwen2.5-32B is most context-sensitive. This suggests that the latter adapts more flexibly to different contexts rather than remaining narrowly focused, potentially due to its exposure to more diverse patterns during training. Overall, scale appears to affect cross-context stability more than architectural differences alone (consistent with scaling-law effects \cite{kaplan2020scaling}).

\subsection{RQ3: Alignment with Category-Mapped Metrics}
Under category-specific prompts, LLM ranking preferences correlate with the corresponding empirical metrics (\autoref{tab:rq3_feature_correlations_sorted}). The strongest and most consistent alignment emerges for fundamental measures, particularly free cash flow, which correlates positively across all model variants. This indicates that, when explicitly asked to focus on a given dimension, LLMs can partially ground their preferences in the underlying firm-level data relevant to that category. Technical indicators such as average trading volume and moving average ratios also show consistent positive associations, suggesting that LLMs are responsive to patterns in market activity when guided by context-specific prompts. Correlations with risk-related metrics are generally negative, reflecting the fact that lower values (i.e., lower risk) correspond to higher LLM confidence rankings.

Despite these findings, the strength and breadth of correlations vary across models, highlighting the influence of both model scale and architecture. Larger models do not uniformly outperform smaller ones, suggesting that architectural refinement and domain-specific representation learning may be as important as scale in improving the alignment between LLM confidence and empirical financial performance. These results indicate partial grounding of LLM judgments in real-world financial patterns, but also leave room for improvement through targeted training and bias mitigation strategies.

    







\begin{table}[ht]
\centering
\caption{Alignment with Category-Mapped Metrics}
\label{tab:rq3_feature_correlations_sorted}
\resizebox{\textwidth}{!}{%
\begin{tabular}{@{}lrrr@{}}
\toprule
\textbf{Feature} & \textbf{Pearson $r$ [95\% CI]} & \textbf{Spearman $\rho$ [95\% CI]} & \textbf{Kendall $\tau$ [95\% CI]} \\
\midrule

\multicolumn{4}{@{}l}{\textbf{Qwen3-32B}} \\
\multicolumn{4}{@{}l}{\textit{Fundamental}} \\
free\_cash\_flow   & 0.568*** [0.405, 0.696] & 0.463*** [0.261, 0.632] & 0.332*** [0.184, 0.470] \\
roa                & 0.318*   [0.130, 0.484] & 0.169     [-0.043, 0.368] & 0.118     [-0.035, 0.259] \\
\multicolumn{4}{@{}l}{\textit{Growth}} \\
earnings\_growth   & 0.023    [-0.180, 0.225] & 0.264*    [0.070, 0.438] & 0.165     [0.034, 0.294] \\
\multicolumn{4}{@{}l}{\textit{Risk}} \\
beta               & -0.252*  [-0.427, -0.058] & -0.201   [-0.378, -0.032] & -0.134   [-0.257, -0.015] \\
volatility\_1y     & -0.313*  [-0.480, -0.124] & -0.229   [-0.404, -0.029] & -0.158   [-0.281, -0.026] \\
volatility\_3m     & -0.328*  [-0.493, -0.140] & -0.250*  [-0.424, -0.060] & -0.163   [-0.291, -0.028] \\
volatility\_6m     & -0.322*  [-0.487, -0.134] & -0.250*  [-0.427, -0.039] & -0.166   [-0.297, -0.026] \\
\multicolumn{4}{@{}l}{\textit{Technical}} \\
avg\_volume\_3m    & 0.394**  [0.214, 0.548] & 0.288*    [0.086, 0.473] & 0.195*    [0.058, 0.341] \\

\multicolumn{4}{@{}l}{\textbf{Qwen3-8B}} \\
\multicolumn{4}{@{}l}{\textit{Fundamental}} \\
free\_cash\_flow   & 0.527*** [0.355, 0.665] & 0.268     [0.048, 0.468] & 0.188*    [0.047, 0.342] \\
roa                & 0.341**  [0.155, 0.504] & 0.221     [0.015, 0.395] & 0.151     [0.023, 0.292] \\
\multicolumn{4}{@{}l}{\textit{Growth}} \\
earnings\_growth   & 0.098    [-0.107, 0.294] & 0.266*    [0.071, 0.445] & 0.173     [0.045, 0.298] \\
\multicolumn{4}{@{}l}{\textit{Technical}} \\
avg\_volume\_3m           & 0.424*** [0.249, 0.573] & 0.370**   [0.164, 0.541] & 0.266**   [0.122, 0.413] \\
moving\_avg\_200d\_ratio  & 0.251*   [0.057, 0.426] & 0.143     [-0.071, 0.347] & 0.083     [-0.055, 0.219] \\
moving\_avg\_50d\_ratio   & 0.280*   [0.089, 0.452] & 0.163     [-0.033, 0.370] & 0.108     [-0.033, 0.247] \\
returns\_1y               & 0.276*   [0.084, 0.448] & 0.122     [-0.088, 0.315] & 0.085     [-0.064, 0.230] \\

\multicolumn{4}{@{}l}{\textbf{Qwen2.5-32B}} \\
\multicolumn{4}{@{}l}{\textit{Fundamental}} \\
free\_cash\_flow   & 0.519*** [0.345, 0.658] & 0.464***  [0.266, 0.628] & 0.323***  [0.168, 0.469] \\
gross\_margin      & 0.355**  [0.171, 0.516] & 0.347**   [0.161, 0.497] & 0.228*    [0.119, 0.334] \\
quick\_ratio       & 0.258    [0.054, 0.442] & 0.286*    [0.097, 0.470] & 0.198*    [0.067, 0.323] \\
\multicolumn{4}{@{}l}{\textit{Growth}} \\
earnings\_growth   & 0.095    [-0.110, 0.292] & 0.287*    [0.084, 0.482] & 0.201*    [0.065, 0.354] \\
\multicolumn{4}{@{}l}{\textit{Risk}} \\
beta               & -0.313*  [-0.480, -0.124] & -0.202  [-0.392, -0.010] & -0.137   [-0.270, 0.002] \\
\multicolumn{4}{@{}l}{\textit{Technical}} \\
avg\_volume\_3m           & 0.382**  [0.201, 0.538] & 0.369**  [0.156, 0.560] & 0.265**  [0.122, 0.407] \\
moving\_avg\_200d\_ratio  & 0.305*   [0.115, 0.473] & 0.204    [-0.007, 0.389] & 0.136    [-0.012, 0.280] \\
moving\_avg\_50d\_ratio   & 0.330*   [0.143, 0.495] & 0.258*   [0.043, 0.462] & 0.180*   [0.026, 0.342] \\
returns\_1y               & 0.301*   [0.112, 0.470] & 0.191    [-0.012, 0.397] & 0.134    [-0.013, 0.279] \\
returns\_3m               & 0.321*   [0.133, 0.486] & 0.216    [-0.006, 0.423] & 0.144    [-0.022, 0.291] \\
\multicolumn{4}{@{}l}{\textbf{Qwen2.5-7B}} \\
\multicolumn{4}{@{}l}{\textit{Fundamental}} \\
current\_ratio     & 0.241    [0.036, 0.427] & 0.274*  [0.065, 0.479] & 0.186*  [0.048, 0.317] \\
free\_cash\_flow   & 0.495*** [0.316, 0.640] & 0.387** [0.171, 0.576] & 0.282** [0.126, 0.431] \\
gross\_margin      & 0.392**  [0.211, 0.546] & 0.392** [0.228, 0.543] & 0.259** [0.152, 0.364] \\
quick\_ratio       & 0.325*   [0.126, 0.499] & 0.330*  [0.135, 0.497] & 0.226*  [0.093, 0.353] \\
roe                & 0.127    [-0.084, 0.328] & 0.289*  [0.084, 0.487] & 0.200*  [0.045, 0.341] \\
\multicolumn{4}{@{}l}{\textit{Growth}} \\
earnings\_growth   & -0.047   [-0.247, 0.157] & 0.287*  [0.086, 0.461] & 0.198*  [0.053, 0.329] \\
revenue\_growth    & 0.204    [0.007, 0.386] & 0.268*  [0.055, 0.442] & 0.171*  [0.017, 0.302] \\
sustainable\_growth\_rate & 0.164 [-0.047, 0.361] & 0.279* [0.071, 0.475] & 0.185* [0.032, 0.320] \\
\multicolumn{4}{@{}l}{\textit{Risk}} \\
sharpe\_ratio      & 0.319*   [0.131, 0.485] & 0.282*  [0.102, 0.449] & 0.185*  [0.053, 0.314] \\
\multicolumn{4}{@{}l}{\textit{Technical}} \\
avg\_volume\_3m           & 0.265*   [0.073, 0.439] & 0.303*  [0.081, 0.503] & 0.207*  [0.053, 0.347] \\
moving\_avg\_200d\_ratio  & 0.261*   [0.069, 0.436] & 0.245   [0.034, 0.434] & 0.163   [0.018, 0.300] \\
returns\_1y               & 0.281*   [0.089, 0.452] & 0.232   [0.023, 0.426] & 0.156   [0.003, 0.289] \\

\bottomrule
\end{tabular}%
}
\footnotesize
\textit{Notes:} *** $p < 0.001$, ** $p < 0.01$, * $p < 0.05$ (FDR corrected). Pearson CI: Fisher's z-transformation; Spearman/Kendall CI: Bootstrap (1000 iterations).
\end{table}

\end{document}